\documentclass[10pt]{article}
\usepackage[letterpaper]{geometry}
\usepackage{hicss}
\usepackage{times}
\usepackage[none]{hyphenat}
\usepackage{url}
\usepackage{latexsym}
\usepackage{mathptmx}
\usepackage{amsmath}
\usepackage{amssymb}
\usepackage{csquotes}
\usepackage{algorithm}
\usepackage{algpseudocode}
\usepackage{booktabs}

\usepackage[
backend=biber,
style=ieee,
sorting=none,
doi=false,isbn=false,url=false 
]{biblatex}

\usepackage{array}

\usepackage{graphicx}

\title{Enhancing Scalability of Optimal Kron-based Reduction of Networks (Opti-KRON) via Decomposition with Community Detection
	\thanks{~This work was supported by the U.S. Department of Energy's Office of Energy Efficiency and Renewable Energy under the Enabling Place-Based Renewable Power Generation using Community Energyshed Design initiative, award number DE-EE0010407. The opinions expressed in this material are those of the authors and do not necessarily reflect those of the U.S. Department of Energy or the United States Government.}
}

\author{Omid Mokhtari\\
	University of Vermont\\
	{\underline{omid.mokhtari@uvm.edu}}  \And
	Samuel Chevalier \\
	University of Vermont \\
	{\underline{samuel.chevalier@uvm.edu} }  \And
	Mads R. Almassalkhi \\
	University of Vermont \\
	{\underline{malmassa@uvm.edu} } }

\date{}

\begin{document}
	\maketitle
	\begin{abstract}
		Electrical networks contain thousands of interconnected nodes and edges, which leads to computational challenges in some power system studies. To address these challenges, we contend that network reductions can serve as a framework to enable scalable computing in power systems. By building upon a prior AC ``Opti-KRON" formulation, this paper presents a DC power flow formulation for finding network reductions that are optimal within the context of large transmission analysis. Opti-KRON previously formulated optimal Kron-based network reductions as a mixed integer linear program (MILP), where the number of binary variables scaled with the number of nodes.  To improve scalability of the Opti-KRON approach, we augment the MILP formulation with a community detection (CD) technique that segments a large network into smaller, disjoint, but contiguous sub-graphs (i.e., communities). For each sub-graph, we then (in parallel) apply MILP-based along with a new cutting plane constraint, thus, enhancing scalability. Ultimately, the new DC-based Opti-KRON method can achieve a 80-95\% reduction of networks (in terms of nodes) while statistically outperforming other CD- and Kron-based methods. We present simulation results for the IEEE RTS-96 and the 2383-bus Polish networks. 
		
	\end{abstract}
	\subsubsection*{Keywords:}
	
	Optimal network reduction, Kron reduction, community detection, mixed integer linear programming (MILP), cutting plane.
	
	\section{Introduction}
	The growing complexity of power grids, driven by the interconnection of diverse regions and the integration of new components, presents a significant challenge for power system studies. Analyzing large-scale power networks, particularly for planning and operation purposes with numerous scenarios, often requires simulating a vast number of potential future states. Unfortunately, the computational demands of these studies can quickly become intractable.
	Network reduction in power grids involves simplifying complex electrical networks by reducing the overall network size while preserving essential characteristics.  
	
	Network reduction methods can be categorized into two primary groups: bus elimination and bus aggregation. The bus aggregation method clusters buses that share similar characteristics, such as buses with similar power transfer distribution factor (PTDF) vectors. These vectors indicate that injections at these buses have similar effects on branch flows. Thus, the generation and load at buses that are considered ``similar" can be aggregated into a single bus~\cite{1525117}. For example,~\cite{10318645} uses a Minimum Spanning Tree (MST) algorithm to identify zones, where within identified zones, nodes are aggregated, particularly focusing on generation bus aggregation.~\cite{oh_new_2010} uses eigenvalue decomposition to find similar components in the PTDF matrix. The proposed model in~\cite{fortenbacher_transmission_2018} depends on pre-determined zones, where the network is divided into zones, and the buses within each zone are aggregated.  After obtaining the reduced PTDF, the susceptances are determined using a nonlinear optimization approach. In~\cite{shi_novel_2015}, bus aggregation is applied to internal zones that are defined based on a set of pre-determined salient lines. Then, the reduced admittance matrix is calculated through a quadratic optimization problem where the objective is minimizing the line flow errors with respect to the full network.
	
	Bus elimination refers to the process of identifying the buses that are kept in the full system and eliminating the rest using either Ward~\cite{5059947} or Kron reduction~\cite{6316101}. Next, the capacities of equivalent branches are determined to create a reduced system that has transfer capacities similar to the full network. For example, the methodology in~\cite{ploussard_efficient_2018} aims to identify critical pairs of buses that likely represent congested lines or potential new line installations. The network is partitioned using the multi-cut problem formulation, which aims to minimize the number of inter-cluster lines by ensuring critical pairs of buses end up in separate clusters. After partitioning, non-border buses are eliminated within each cluster using Kron reduction.~\cite{10122726} proposes a Ward-based reduction model that incorporates nonlinear load functions to estimate AC power flow across a range of operating points. Using a mixed integer optimization method, the parameters of the reduced DC model, including the topology of the external equivalent network and the allocation of power injections at boundary buses, are obtained. The framework in~\cite{8017469} is adaptable to generate either Ward reduction, bus-aggregation-based reduction, or a hybrid of both, where it treats the Ward reduction as an optimization problem, aiming to preserve the PTDF matrix's submatrix that relates to the retained buses and branches.
	
	These methods, however, have used predefined sets of nodes, zones, or lines to remain in the grid. The problem of which set of nodes and how many nodes should remain in the reduced network have been solved by the Opti-KRON model in \cite{9992730}, where an optimization problem is designed to find the optimal network reduction with minimum voltage deviation from the full network. Even though based on Kron reduction, this approach can be considered a hybrid implementation of bus aggregation and bus elimination, where each eliminated node is assigned to a super-node that remains. As a continuation of the methods proposed in~\cite{9992730}, our paper enhances the scalability of the approach, and it explores larger instances of optimal network reduction.
	
	To enhance network reduction scalability, strategic network partitioning is employed, with community detection serving as a crucial method to effectively partition the network into distinct, manageable segments. Community detection is a technique used in network science to identify groups or communities within a network where nodes are more densely connected to each other than to nodes in other communities \cite{girvan2002community}. The main goal of community detection is to find a structure within a large network that helps in understanding the nature of the connections and interactions among its components. The advantage of community detection over other network partitioning approaches is that it does not require a predefined set of nodes or lines for network decomposition; instead, the network is decomposed based on the natural topological features of the network.

	Building upon the previous Opti-KRON model, this work makes several key contributions to the field of network reduction:
	\begin{itemize}
		\item Proposing a DC Framework for the Opti-KRON Model: We change the formulation of Opti-KRON from AC to DC power flow in order to extend applicability to transmission networks and simplify the computations.
		\item Embedding Community Detection with Network Reduction: We enhance scalability by integrating community detection techniques with network reduction processes, enabling more efficient handling of large-scale networks.
		\item Using a Cutting Plane and Big-M Linearization: We further improve scalability by employing a cutting plane restriction and Big-M linearization, which accelerates the reduction process.
	\end{itemize}
	The remaining paper is structured as follows. Section \ref{sec:model}
	summarize the network model and Kron reduction. In Section \ref{sec: formulation}, the MINLP formulation for Kron-based
	network reduction is presented. Reformulation, scalability, and community detection are presented in section \ref{sec: scalability}. Experimental results are proposed in section \ref{sec: results}. Finally, the paper concludes in section \ref{sec: end} with a summary and a brief discussion on future directions and applications.
	
	\section{Network model and Kron reduction}\label{sec:model}
	
	\subsection{Network Representation}
	We model a power system network as a graph $\mathcal{G} = (\mathcal{V}, \mathcal{E})$ with $\mathcal{V}$ and $\mathcal{E}$ representing vertices and edges, where $|\mathcal{V}| = n$ and $|\mathcal{E}| = m$. The network's structure is defined by the adjacency matrix ${\Lambda} \in \mathbb{R}^{n \times n}$, where $\Lambda_{ij}$ indicates the presence or absence of edges, corresponding to transmission lines. Additionally, the complex nodal admittance matrix, or the $Y$-bus matrix $\mathbf{Y}_b \in \mathbb{C}^{n \times n}$, is utilized to capture the electrical characteristics of the network.
	
	\subsection{DC Power Flow Model}
	DC power flow is an approximate model for calculating power flow in an electrical network by assuming negligible resistance, small phase angle differences, and constant voltage magnitudes, resulting in a linear relationship between power flow and phase angle differences. 
	The power flow equation is represented as $P = B \boldsymbol{\theta}$, where $P$ is the net power injection vector, $\boldsymbol{\theta}$ the node phase angles, and $B={\rm Im}\{\mathbf{Y}_b\}$ the susceptance matrix, derived from the $Y$-bus matrix.
	
	\subsection{Kron Reduction}
	The Kron reduction of a given graph yields another graph, where the admittance matrix is derived by taking the Schur complement of the original admittance matrix with respect to a specified subset of nodes. Thus, a kron-reduced equivalent of a network with admittance matrix $\mathbf{Y}_b$, given a set of nodes to be kept ($k$) and a set of nodes to be removed ($r$), can be calculated as follows:
	\begin{align}
		Y^{\rm kron} = Y_{kk} - Y_{kr} {Y_{rr}}^{-1} Y_{rk} 
	\end{align}
	Additionally, instead of calculating the Kron-reduced admittance matrix, the equivalent impedance matrix of a Kron-reduced network can be directly calculated by removing the rows and columns corresponding to the eliminated nodes from the impedance matrix of the full network \cite{9992730}. This approach facilitates the direct calculation of power flow for the Kron-reduced network which has been addressed in the next section.
	
	\section{Proposed Model}\label{sec: formulation}
	According to \cite{9992730}, in a network reduction problem, in AC environment, the voltage of the kept nodes ($V$) can be calculated through:
	\begin{equation}\label{eq: opti_kron_voltage}
		V = \text{diag}(s) Z_b A I,
	\end{equation}
	where $s$ is the vector of binary decision variables that selects the nodes to be kept in the network. When $s_i = 1$ node $i$ remains in the network as a \textbf{super-node}, and when $s_i = 0$ node $i$ will be eliminated. $Z_b$ is the impedance matrix, and $A$ is the matrix binary variable which codifies where currents from reduced nodes are placed i.e. when $A^{i,j} = 1$ injected current of node $j$ will be added to node $i$. Vector $I$ is the current injection at each node.
	
	While in a DC environment, $|V| = 1$ and the admittance matrix is singular, so equation \eqref{eq: opti_kron_voltage} cannot be directly used. However, the equivalent of \eqref{eq: opti_kron_voltage} in a DC power flow context can be expressed as
	\begin{subequations}
		\begin{align}
			&\theta = \text{diag}(s) \Psi \\
			&B \Psi = A P,
		\end{align}
	\end{subequations}
	where parameter $P$ is the real power injection at each node, $\theta$ is the DC voltage angle of the super-nodes, and $B$ is the imaginary part of the admittance matrix. $\Psi$ is the auxiliary variable vector that contains voltage angles of all nodes.
	
	With this formulation, each reduced node belongs to a cluster with one super-node as the representative of that cluster where the power injection of all nodes in each cluster is assigned to their super-nodes.
	
	In this paper, our goal is to identify nodes to retain in such a way that the reduced network accurately represents the original grid. Therefore, we need to define an objective function that, while encouraging a greater level of reduction, should prevent deviation from the original grid. In order to accomplish this, Maximum Intra-Cluster Error (MICE) is defined to calculate the maximum error in each cluster. In this paper, the error of a node is defined as the difference between its voltage angle in the original grid (obtained from DC power flow solutions on the original network) with the voltage angle of its super-node in the cluster it belongs to (obtained from performing power flow solutions on the reduced network). This formulation helps us to minimize the maximum error in each cluster. Thus, the objective function that optimally balances the trade-off between reduction levels and the deviation can be expressed as
	\begin{align}\label{eq: obj1}
		\mathcal{L}={ \sum_{c} \left\Vert {\hat{\theta}}_c - \Theta_c \right\Vert_{\infty}} - \alpha \sum_{i=1}^n(1-s_{i}),
	\end{align}
	where the parameter $\alpha$ is a weighting factor that can be used to determine the network reduction priority e.g., higher values of $\alpha$ lead to a greater level of reduction with higher error while a smaller $\alpha$ preserves accuracy. $\theta_c$ is the given data vector of voltage angles, and variable $\Theta_c$ is the cluster angle of the reduced network. However, to calculate voltage angles and distinguish them based on their clusters, we need to reformulate the objective function and utilize the optimization problem below:
	\begin{subequations}
		\begin{align}
			\min_{\mathbb{E}} \quad &{ \sum_{i}^n  \overbrace{ \left\Vert \vec{e}_i^T X \right\Vert_{\infty}}^{\text{MICE}} } - \alpha  \sum_{i}^n(1-s_{i}) \label{eq: obj2}\\
			{\rm s.t.}\;\; \quad &X = A \text{ diag}(\hat{\theta} - \Theta) \label{error}\\
			&\theta = \text{diag}(s) \Psi \label{eq: thet_psi}\\
			&B \Psi = A P\label{eq: zap}\\
			&\Theta = A^T \theta \label{eq: tat}\\
			&\mathbf{1}^T A = \mathbf{1}^T \label{eq: assign_limit1}\\
			&A \leq s \mathbf{1} \label{eq: assign_limit2}\\
			& A_{i,i} = s_i,\ \forall i\ \in N \label{eq: diagA}\\
			&A_{i,j} \leq \Lambda_{i,j},\ \forall i,j\ \in N, i\neq j,\label{eq: assign_adj}
		\end{align}
	\end{subequations}
	where  $\mathbb{E} :=\{ X, \Psi, \theta,\Theta, s, A\}$, $n$ is the number of nodes in the full grid, $N$ is the set of all nodes, and $\vec{e}_i$ is the $i$th natural basis vector, where all elements are zero except for the $i$-th element. Equation~(\ref{error}) calculates the error of all nodes in the clusters. If node $i$ is a reduced node, row $i$ of $X$ is zero, and if node $i$ is a super-node, row $i$ represents errors of that cluster. Thus, MICE for each cluster is equal to the maximum value of each row of matrix $X$. The voltage angles of super-nodes are determined through \eqref{eq: thet_psi} and \eqref{eq: zap}. Equation (\ref{eq: tat}) assigns voltage angles of super nodes to the children nodes within their clusters. Each node can only be assigned to one cluster, forced by \eqref{eq: assign_limit1}. According to \eqref{eq: assign_limit2}, reduced nodes can be assigned only to super-nodes. Each non-reduced bus keeps its own current by equation \eqref{eq: diagA}. Node assignment is restricted to neighbor nodes in \eqref{eq: assign_adj}, where $\Lambda$ is the adjacency matrix.
	
	Since \eqref{eq: assign_adj} limits the degree of reduction, we need to employ an iterative algorithm to achieve smaller networks. In the algorithm introduced next, we use Kron-reduction to update the $Y$-bus and then the adjacency matrix of the reduced network. However, to calculate MICE for all the nodes in the original grid, it is necessary to keep track of child-nodes at each iteration. Thus, we define relations
	\begin{subequations} \label{eq: 7}
		\begin{align}
			\Omega^k &= A^k \Omega^{k-1} \label{eq: A_supmat}\\
			s^k &\leq s^{k-1}, \label{eq: S_supvec}
		\end{align}
	\end{subequations}
	where $k$ denotes the iteration index and $\Omega$ is the matrix containing the cluster information. While matrix $A$ has only the information of the current iteration (no information about the reduced nodes in the previous iterations), we track previously reduced nodes through $\Omega$. Specifically, element $\Omega_{ij} = 1$ if reduced node $j$ is assigned to super-node $i$.  
	
	For the initial iteration, the setup is as follows: $s^0 = \mathbf{1}_n$, indicating that every node is considered as a potential super-node. Concurrently, $\Omega^0$ is set to the identity matrix $I$. As the algorithm progresses, these matrices are updated to reflect the evolving cluster structures and super-node assignments. 
	
	In addition to \eqref{eq: A_supmat} and  \eqref{eq: S_supvec}, equations (\ref{eq: tat}) and (\ref{eq: assign_limit1}) need to be altered as follows:
	\begin{subequations} \label{eq: 8}
		\begin{align}
			\Theta^k &= (\Omega^k)^T  \theta^k \label{eq:                     Thet_supmat}\\
			\mathbf{1}_n^T A^k &= (s^{k-1})^T. \label{eq: A_supvec}
		\end{align}
	\end{subequations}
	Thus, the new optimization problem can be expressed as:
	\begin{subequations} \label{eq:final_nlp}
		\begin{align} 
			\min_{\mathbb{E}^k} \quad &{ \sum_{i}^n  \overbrace{ \left\Vert \vec{e}_i^T X^k \right\Vert_{\infty}}^{\text{MICE}} } - \alpha  \sum_{i}^n(1-s_{i}^k) \label{eq: obj3}\\
			&X^k = \Omega^k \text{ diag}(\hat{\theta} - \Theta^k) \label{error2}\\
			&(\ref{eq: thet_psi}),(\ref{eq: zap}),(\ref{eq: assign_limit2}), \eqref{eq: diagA}, \eqref{eq: assign_adj}, \eqref{eq: 7},  \eqref{eq: 8}
		\end{align}
	\end{subequations}
	Where $\mathbb{E}^k =\{ X^k, \Psi^k, \theta^k,\Theta^k, \Omega^k, s^k, A^k\}$. Problem (\ref{eq:final_nlp}) is an NP-hard Mixed-Integer Nonlinear Programming (MINLP) problem, which poses a computational challenges. 
	In the following section, we will improve scalability of this formulation.
	
	\section{Scalability}\label{sec: scalability}
	To enhance scalability of the proposed formulation, we first employ Big-M reformulation for the bi/non-linear expressions. Secondly, a cutting plane is added to the formulation to reduce space of feasible super-node assignments. Finally, community detection techniques are applied to segment large networks into smaller, disjoint, but contiguous sub-graphs (i.e., communities), which enables us to parallelize the optimization problem for each community. These techniques are explained in detail in the following sections.
	
	\subsection{Big-M Reformulation}
	To reformulate non-linear (\ref{eq: thet_psi}) and (\ref{eq: Thet_supmat}) as a linear expression, we employ the Big-M method. The equivalent set of inequality constraints  for (\ref{eq: thet_psi}) is as follows:
	\begin{subequations} \label{eq: lin1}
		\begin{align}
			&	\theta \leq (\mathbf{1}-s)M + \Psi \\
			&	\theta \leq s M \\
			&	\theta \geq (s-\mathbf{1})M + \Psi\\
			&	\theta \geq -sM.
		\end{align}
	\end{subequations}
	To linearize equation (\ref{eq: Thet_supmat}), we need to define an auxilary row vector $\gamma_i$ and a selection vector $\vec{e}_i$ such that:
	\begin{subequations} \label{lin_2}
		\begin{align}
			&\gamma_i={\vec{e}}_i^T\left({\Omega}\right)^T,\ \forall i\ \in N\\
			&\varphi_i\le\left(\mathbf{1}^T-\gamma_i\right)M+\theta^T,\ \forall i\ \in N\\
			&\varphi_i\le\gamma_iM,\ \forall i\ \in N\\
			&\varphi_i\geq\left(\gamma_i-\mathbf{1}^T\right)M+\theta^T,\ \forall i\ \in N\\
			&\varphi_i\geq-\gamma_iM,\ \forall i\ \in N\\
			&\Theta = \varphi \mathbf{1},\label{eq: Thet_phi}
		\end{align}
	\end{subequations}
	where $\varphi_i$ denotes row $i$ of $\varphi$.
	Equation \eqref{error2} can also be linearized in the same way: 
	\begin{subequations} \label{lin_3}
		\begin{align}
			&l = (\hat{\theta} - \Theta) \\
			&\zeta_i = \vec{e}_i^T \Omega,\  \forall i\ \in N \\
			&X_i \le \left(\mathbf{1}^T - \zeta_i\right)M+ l^T,\ \forall i\ \in N     \\
			&X_i \le \zeta_i M,\  \forall i\ \in N     \\
			&X_i \geq (\zeta_i - \mathbf{1}^T)M + l^T,\  \forall i\ \in N\\
			&X_i \geq - \zeta_i M,\  \forall i\ \in N. 
		\end{align}
	\end{subequations}
	Similarly, $X_i$ denotes row $i$ of matrix $X$.
	As too large values for $M$ can cause numerical issues during solving optimization problems, we need to estimate the smallest values that satisfy the equations. Thus, proper values of $M$ depend on the operating conditions of each network. For this problem, we choose $M$ based on the voltage angles, acquired from DC power flow. However, generally, the ideal value of \( M \) lies in the range \(\frac{\pi}{6} \leq M \leq \frac{\pi}{2}\). 
	
	Another source of nonlinearity arises from the infinity norm in the objective function, which is classically reformulated using inequality constraints. Combining the different Big-M and $\infty$-norm reformulations replaces all non-linearities with the following MILP:
	\begin{subequations} \label{eq: obj_final}
		\begin{align}
			\min_{\mathbb{E}^k} \quad &\sum_{i}^n {\mu_i^k } - \alpha  \sum_{i}^n(1-s_{i}^k) \label{eq: obj4}\\
			-\mu_i^k \leq &\vec{e}_i X^k \leq \mu_i^k ,\ \forall i\ \in N \\
			\begin{split}
				(\ref{eq: thet_psi}),&(\ref{eq: zap}),(\ref{eq: assign_limit2}), \eqref{eq: diagA}, \eqref{eq: assign_adj},\\ \eqref{eq: 7}, &\eqref{eq: A_supvec}, \eqref{eq: lin1},\eqref{lin_2},\eqref{lin_3}.
			\end{split}
		\end{align}
	\end{subequations}
	
	The new formulation of our problem as a MILP in (\ref{eq: obj_final}) has no nonlinear terms. 
	However, a MILP is still NP-hard, which poses computational challenges. This is in part due to a very large search space, which can significantly challenge the convergence rate and computational efficiency of the optimization process. Thus, next, we propose to reduce the size of the feasible set with cutting planes and community detection. %

	\subsection{Cutting Plane}
	The cutting plane simplifies the problem by reducing the feasible set. The following equation acts as such a cutting plane by limiting the number of simultaneous reductions in the network:
	\begin{align}\label{eq: CP}
		\sum_i^{n_s^k} {s_i^k} \geq n_s^k - q.
	\end{align}
	In \eqref{eq: CP}, $n_s^k$ is the number of super-nodes in the network at iteration $k$, and parameter $q$ is the number of nodes allowed to be reduced. Adding this constraint to the MILP problem described earlier, i.e. \eqref{eq: obj_final}, results in a more tractable optimization problem, which we refer to as \textbf{DC Opti-KRON}.
	
	\subsection{Community Detection}
\begin{figure*}[thb]
		\centering
		\includegraphics[width=1\linewidth]{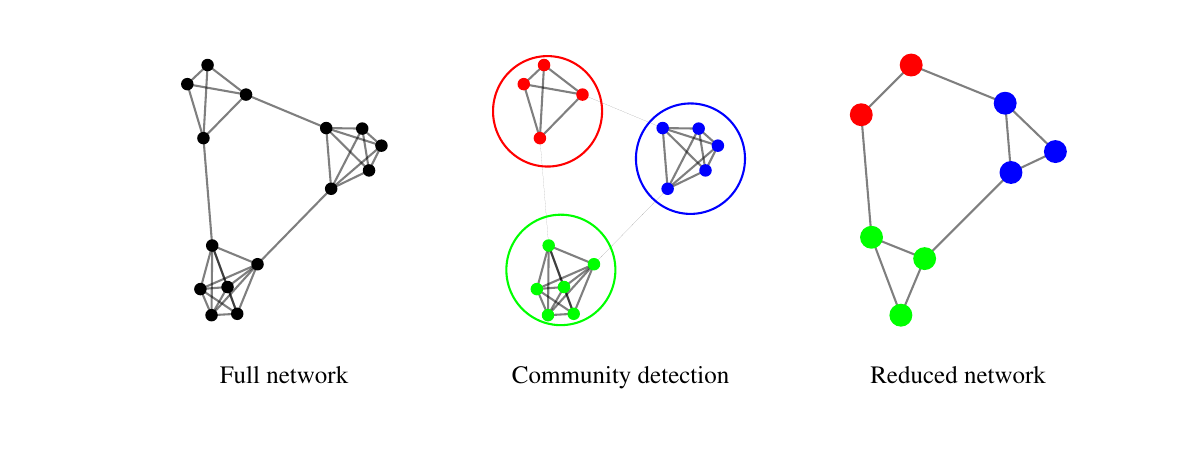}
		\caption{
			Decomposing the full network (left) with community detection to detect each disjoint community (colored sub-graphs in middle), and reconstructing the Kron-reduced network based on applying DC Opti-KRON to each community in parallel (right).}
		\label{fig:problemflow}
	\end{figure*}

	Community detection refers to the process of identifying groups of nodes in a network that are more densely connected internally than with the rest of the network~\cite{girvan2002community}.  
	By leveraging community detection in electrical networks, we can efficiently segment the network into distinct communities. For each community, we then apply DC Opti-KRON in parallel, accelerating the network reduction method. Hence, based on the nodes belonging to each cluster, admittance, adjacency, and susceptance matrices of the communities are made from the full admittance matrix. Additionally, to enforce power balance in each community, power injections of coupling nodes are updated according to the flow of lines that connect communities.  After reducing decoupled communities, We store clusters' information i.e. matrix $\Omega$ and vector $s$, including information for super-nodes and reduced nodes.  We aggregate $s$ vectors and find the final kron reduce network based on that. Additionally, we use $\Omega$ matrices to find new power injections. This method enables us to rebuild a reduced version of the full network from reduced sub-networks. 
	Algorithm~\ref{alg: CBNR} describes the presented optimal network reduction method and is illustrated with Fig.~\ref{fig:problemflow}.
	
	\begin{algorithm}\
		\caption{Community-Detection Based Network Reduction Algorithm}
		\label{alg: CBNR}
		\begin{algorithmic}[1]
			
			\State \textbf{Input Data:} $Y_b$, $P$, $\hat{\theta}$, $\alpha$, $q$
			
			\State \textbf{Community Detection:} Find best possible communities.
			
			\For{each community}
			\State Find $Y_{\rm community}$, $P_{\rm community}$, $\hat{\theta}_{\rm community}$ to use in DC Opti-KRON.
			\State Update: $k = 0$, $s^{0} = \mathbf{1}$, $\Omega^{0} = I$, 
			\While{$n_s^k < n_s^{k-1}$}
			\State $k \gets k + 1$
			\State \textbf{Solve DC Opti-KRON.}
			\State Update $n_s$, $\Omega$, $\Lambda$, $P$.
			\EndWhile
			\State Store final $\Omega$ and $s$ of the reduced community.
			\EndFor
			
			\State Reconstruct reduced network.
			
		\end{algorithmic}
	\end{algorithm}
	
	\section{Experimental Result} \label{sec: results}
	This section presents the simulation results for the proposed model applied to the IEEE 96-RTS test system and 2383-bus Polish transmission network. To solve optimization problems we used Gurobi solver with the optimality gap set to 0. To accurately simulate the varying operational conditions of the network in our model, we incorporated both high- and low-loading scenarios. 
	\subsection{IEEE RTS96 73 bus Test System}
\begin{figure}[thb]
		\centering
		\includegraphics[width=1.05\linewidth]{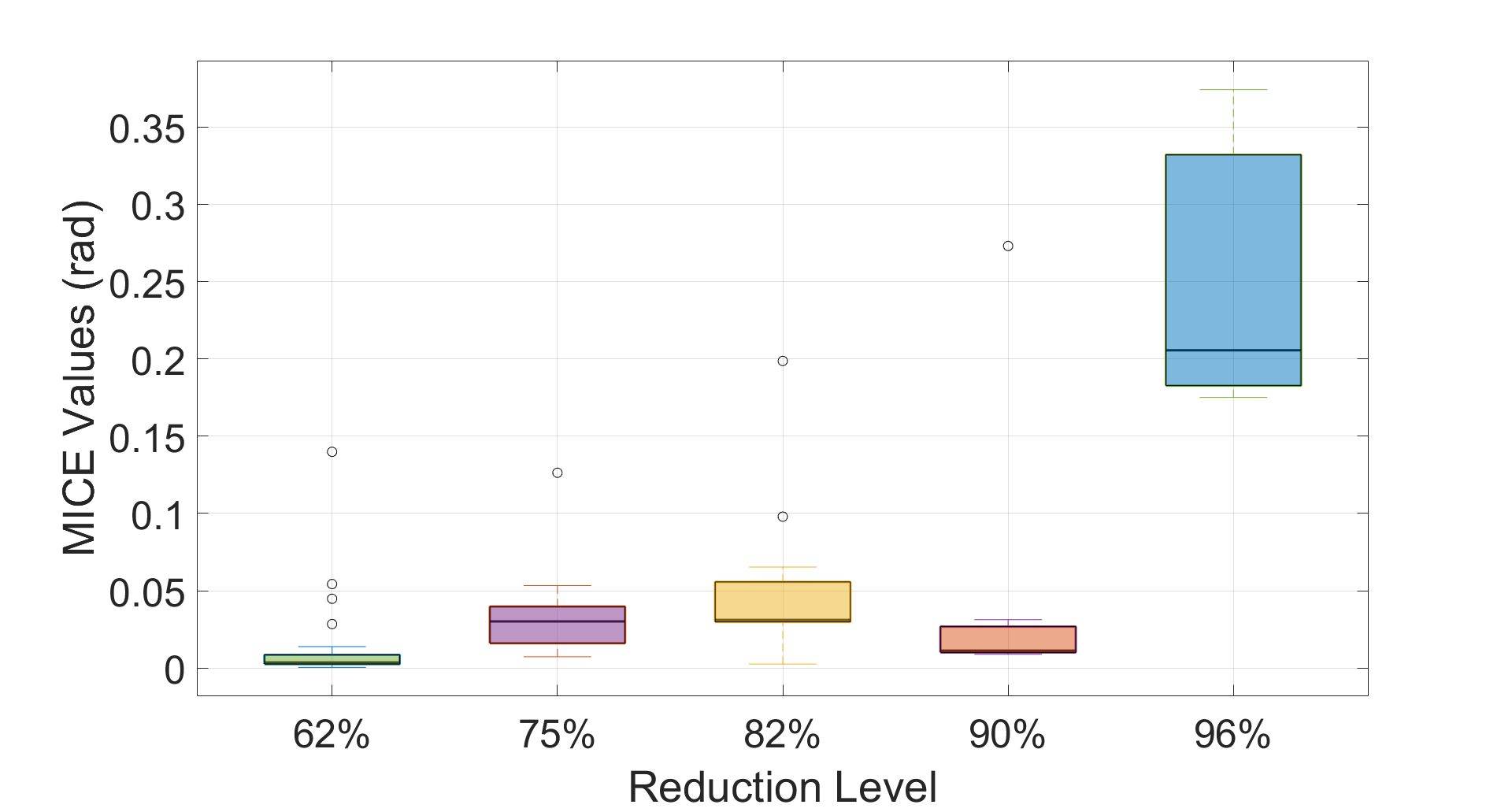}
		\caption{Distribution of Maximum Intra-Cluster Error (MICE) with different reduction levels of RTS96 73 bus system.}
		\label{fig:MICE73}
	\end{figure}

	The IEEE  RTS96 73 bus test system consists of three identical 24-bus systems with an extra bus to connect Area~1 and Area~3. We applied the proposed network reduction model to this system with different values for $\alpha$ to obtain different reduction levels.  Fig. \ref{fig:MICE73} represents the distribution of MICE with different levels of reduction. 
	
	Each data point in Fig. ~\ref{fig:MICE73} represents the maximum error among all nodes that belong to a cluster. For example, for the 96\% reduced network, there are only 3 Super-nodes in the reduced grid and MICE values regarding these Super-nodes are approximately 0.37, 0.2, and 0.17 respectively. Notably, the network with a 62\% reduction level exhibits the lowest MICE values, aligning with expectations that lower reduction levels typically yield lower errors. However, the relationship between reduction level and MICE is not strictly monotonic. For example, the network reduced by 90\% demonstrates lower MICE values compared to those reduced by 75\% and 82\%, despite a higher reduction percentage. This non-monotonic relationship can be attributed to the iterative nature of our algorithm, which follows different reduction paths for each scenario.
	
	To investigate the effect of the cutting plane on the solution time and the accuracy, different values for $q$ have been used in simulations. Table \ref{tab:Cut} summarize the this investigation. Where $q$ equals the total number of nodes in each iteration ($n_s$), signifies no cutting plane is used, which results in a minimal number of iterations (3 iterations). Now, we can categorize values of $q$ into three. For higher $q$ values (4 and 5), iterations increase, and total time extends compared to no cutting plane, but each iteration is significantly quicker. This suggests a trade-off between total duration and speed per iteration. The optimal $q$ value for this network appears to be 3, striking a balance by substantially reducing the time per iteration efficiently while maintaining the reduction integrity. In contrast, lower $q$ values (1 and 2) result in the fastest iteration times but at a significant cost to reduction accuracy, with levels plummeting below optimal (with no cutting plane). This highlights that while lower $q$ values enhance speed, they significantly compromise the quality of network reduction. 
	
	\begin{table}[thb]
		\centering
		\caption{Optimization problem performance with different cutting planes \vskip 3pt}
		\label{tab:Cut}
		\begin{tabular}{>{\centering\arraybackslash}m{1.5 cm}|>{\centering\arraybackslash}m{1cm}|c|>{\centering\arraybackslash}m{1.3cm}|m{0.7 cm}}
			\hline 
			\bf $q$ & \bf Time (s) & \bf Iteration & \bf Avg Time/Iter (s) & \bf Red. Level \\ 
			\hline
			$n_s$ (no CP)& 2703 & 3 & 901.00 & 96\% \\
			5 & 5871 & 15 & 391.40 & 96\% \\
			4 & 4105 & 18 & 228.05 & 96\% \\
			3 & 677 & 24 & 28.21 & 96\% \\
			2 & 485 & 19 & 25.53 & 52\% \\
			1 & 478 & 32 & 14.94 & 44\% \\
			\hline
		\end{tabular}
	\end{table}
	
	\subsection{2383-bus Polish transmission network}
\begin{figure}[thb]
		\centering
		\includegraphics[width=1\linewidth]{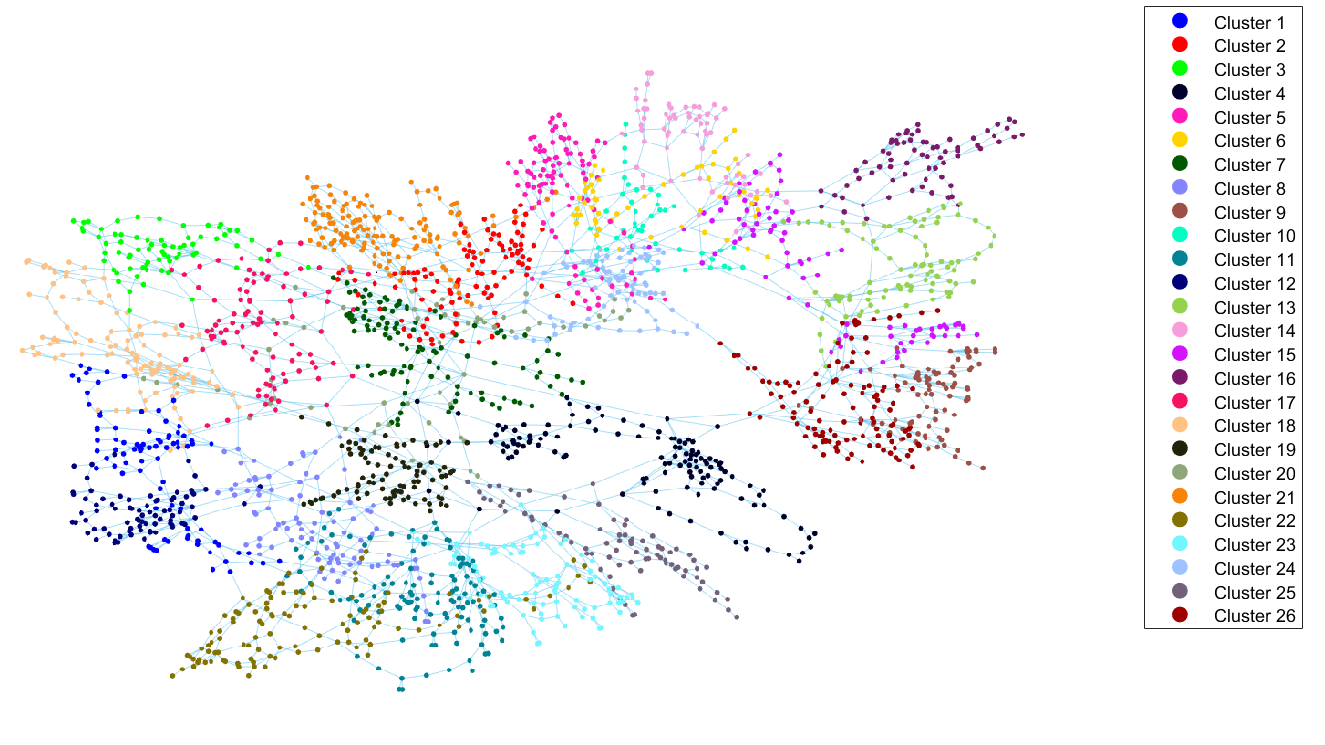}
		\caption{Visualization of the 2383-bus Polish transmission network with community detection.}
		\label{fig: fullnetwork}
	\end{figure}

	To identify communities within this network, we employed 2 approaches: weighted and unweighted. Fast Greedy Maximization Method \cite{clauset2004finding} has been used to find the communities based on the unweighted graph Laplacian, which revealed 26 distinct communities. Figure \ref{fig: fullnetwork} visualizes the network graph, where nodes belonging to the same community are highlighted with unique colors. The detected communities yielded a notably high modularity score of 0.9. This score indicates a strong structure of communities within the network, signifying that the nodes within each detected community are more densely interconnected as compared to nodes in different communities \cite{newman2004finding}. 
	
\begin{figure}[thb]
		\centering
		\includegraphics[width=1\linewidth]{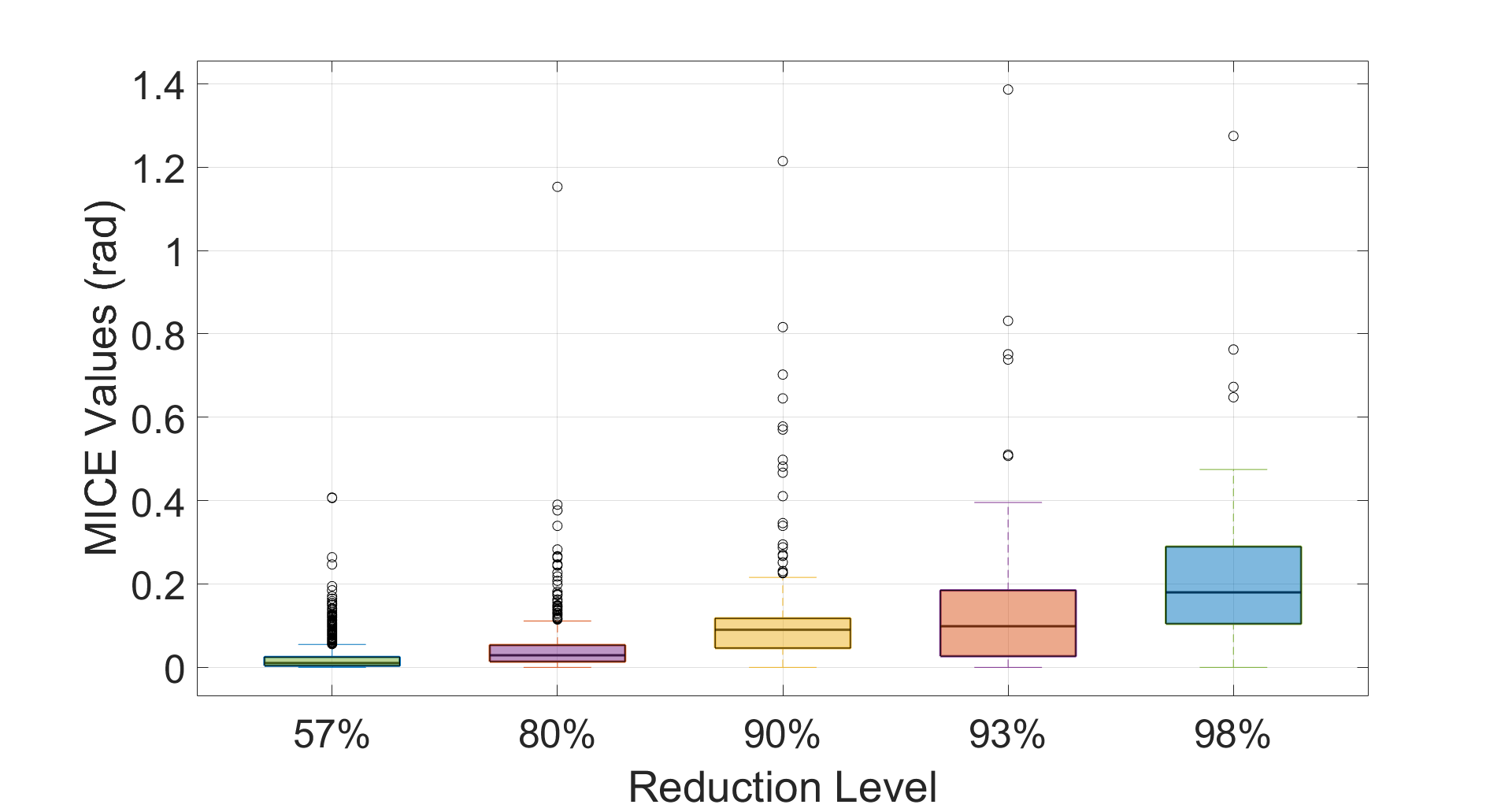}
		\caption{Distribution of MICE with different reduction levels of Polish transmission network with unweighted community detection.}
		\label{fig:Box_pol_unw}
	\end{figure}

	Fig. \ref{fig:Box_pol_unw} represents the distribution of maximum error inside each cluster with different levels of reduction for this network. After completing the network reduction, which includes mapping nodes to super-nodes based on community detection results, we compile these mappings into a pool that defines the reduced network's new structure. This reconstituted network is then subjected to a power flow analysis to assess the accuracy of the reduction. The MICE values are calculated by comparing the voltage angles and other power flow parameters between the original and the reconstructed reduced network, providing a direct measure of the reduction's impact on network performance.
	The presence of outliers in the MICE distribution can be attributed to certain communities within the network that, post-reduction, adopt a radial-like configuration with a limited number of nodes. In these cases, the nature of DC power flow—where voltage angles are particularly sensitive in radial structures—can lead to heightened variability in error measures. To address this, identifying the communities that contribute to these outliers is critical. Subsequently, revisiting the network reduction process for these specific communities with modified reduction constraints could help achieve a more balanced error distribution and enhance the overall effectiveness of the network reduction. 
	
\begin{figure}[thb]
		\centering
		\includegraphics[width=1\linewidth]{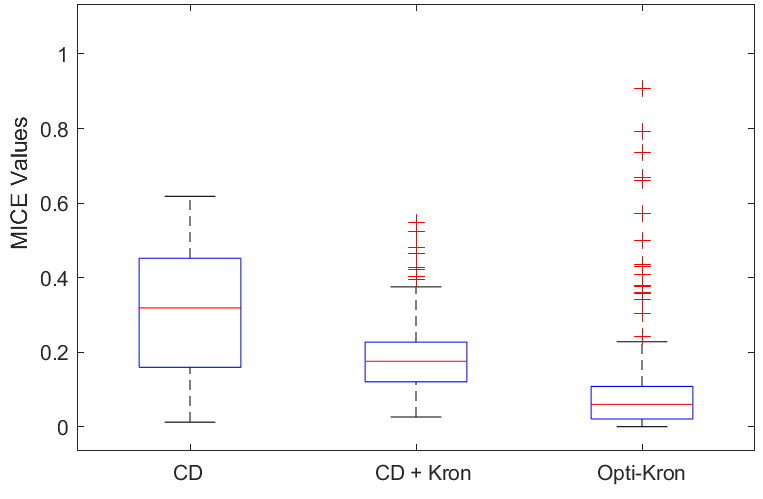}
		\caption{MICE distribution for network reduction of a 90\% reduced Polish network.}
		\label{fig: box_90}
	\end{figure}

	To compare how different network reduction approaches can affect MICE, we also used 2 novel approaches for this purpose. To do so, first, we select the number of nodes we want to keep in the grid (reduction level), and in the second step, we find the same number of communities using community detection. To choose super-nodes in each community as the representative of all the nodes within the communities, central nodes should be selected, where we aggregate all nodes in each community into its central node. Then, interconnection among communities can remain the same as the full network or can be computed through Kron reduction. Fig. \ref{fig: box_90} demonstrates MICE for 90\% reduction level for these approaches compared to DC Opti-KRON. 'CD' represents the error after aggregating communities into single nodes, while 'CD + Kron' applies Kron reduction using the central node of communities. 'Opti-KRON' is our proposed optimization technique coupled with community detection. Even though DC Opti-KRON has some outliers, the average of MICE for this approach is smaller than the others.
	
\begin{figure}[thb]
		\centering
		\includegraphics[width=1.1\linewidth]{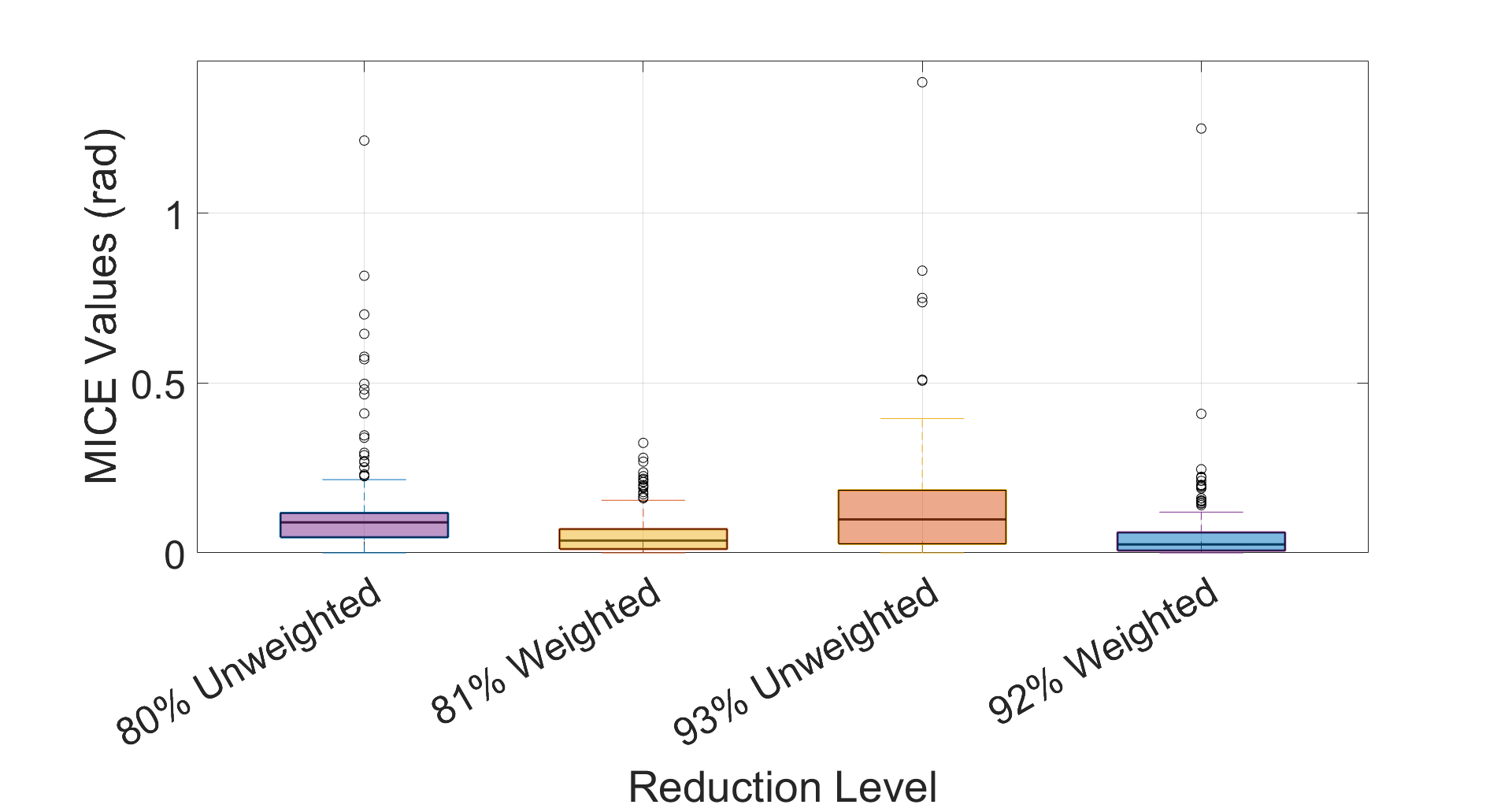}
		\caption{MICE comparison for reduced networks based on community detection with weighted and unweighted graph Laplacian.}
		\label{fig: weighted_mice}
	\end{figure}

	In addition to unweighted graph laplacian, weighted graph laplacian is implemented to be utilized in community detection. In this regard, spectral Clustering has been used, and based on the modularity metric, 26 communities have been selected. Fig. \ref{fig: weighted_mice} represents MICE distribution for approximately 81\%  and 92\% reduced networks using both weighted and unweighted graph Laplacian. The weighted approach incorporates physical (line admittances), which influence power flow, resulting in communities that align more closely with the physical structure of the network. This consideration leads to smaller deviations in MICE and a reduction in both the number and magnitude of outliers compared to the unweighted approach.
	
	While both the weighted and unweighted methods identified 26 communities, they differ significantly in their characteristics. Communities derived from the weighted graph Laplacian are more `physics-aware', reflecting actual power flows and network interactions more accurately. Conversely, the unweighted approach, while potentially less reflective of the physical network properties, leads to communities with fewer coupling nodes. This characteristic can be advantageous as it results in simpler, more defined community boundaries.

	\section{Conclusion and Future work} \label{sec: end}
	This paper has presented a novel approach to enhancing the scalability of network reduction for large transmission networks. By integrating community detection techniques in conjunction with a DC-based Opti-KRON formulation, we have scaled Opti-KRON to optimally reduce a 2000-bus network (past work was limited to networks with 100s of nodes). This is achieved by having community detection segment the network into smaller, coherent sub-networks, on which we can readily (and in parallel) apply Opti-KRON to reduce the overall network. The experimental results on the IEEE 96-RTS test system and the 2383-bus Polish transmission network validate the efficacy of our approach. The DC Opti-KRON model consistently shows lower Maximum Intra-Cluster Error (MICE) values compared to other reduction methods.
	
	Future work could focus on enhancing community detection by incorporating external equivalency feedback, allowing each community to be analyzed independently while still influenced by broader grid dynamics. Another direction for future work is to integrate community detection and the cutting plane methodologies with the AC Opti-KRON formulation, which is suitable for distribution networks. This integration would enable a comprehensive co-reduction method. These advancements could significantly enhance the scalability and applicability of the Opti-KRON model in complex power systems.

	\printbibliography
	
\end{document}